\documentclass[10pt,usletter,conference]{IEEEtran}
\usepackage[top=2.5cm, bottom=2.0cm, left=1.45cm, right=1.45cm]{geometry}

\usepackage{graphicx}
\usepackage{subfigure} 
\usepackage{multirow}
\usepackage{makecell}
\usepackage[absolute,showboxes]{textpos}
\usepackage{hyperref}

\TPMargin{5pt}

\newcommand{\copyrightstatement}{
    \begin{textblock}{0.84}(0.08,0.95)    
         \noindent
         \footnotesize
         \copyright 2020 IEEE. Personal use of this material is permitted. Permission from IEEE must be obtained for all other uses, in any current or future media, including reprinting/republishing this material for advertising or promotional purposes, creating new collective works, for resale or redistribution to servers or lists, or reuse of any copyrighted component of this work in other works.
         DOI: \href{https://ieeexplore.ieee.org/abstract/document/9371923}{10.1109/IEDM13553.2020.9371923}
    \end{textblock}
}

\begin{document}

\copyrightstatement

\title{\huge \textit{Ab initio} simulation framework for Majorana
  transport in 2D materials: towards topological quantum computing}

\author{\authorblockN{%
Y.~Lee, T.~Agarwal, and M.~Luisier}
\authorblockA{Integrated Systems Laboratory, ETH Zurich, Zurich,
  Switzerland, email: youslee@iis.ee.ethz.ch}}

\maketitle

\begin{abstract}
  We present an \textit{ab initio} modeling framework to simulate
  Majorana transport in 2D semiconducting materials, paving the way
  for topological qubits based on 2D nanoribbons. By combining
  density-functional-theory and quantum transport calculations, we
  show that the signature of Majorana bound states (MBSs) can be found
  in 2D material systems as zero-energy modes with peaks in the local
  density-of-states. The influence of spin-orbit coupling and external
  magnetic fields on Majorana transport is studied for two relevant 2D
  materials, WSe$_2$ and PbI$_2$. To illustrate the capabilities of
  the proposed \textit{ab initio} platform, a device structure capable
  of hosting MBSs is created from a PbI$_2$ nanoribbon and carefully
  investigated. These results are compared to InSb nanowires and used
  to provide design guidelines for 2D topological qubits.
\end{abstract}

\section{Introduction}
\par Quantum computer promises an exponentially augmented computing
power to solve specific problems that are almost impossible to tackle
using classical machines based on the von Neumann architecture
\cite{Feynman}. Because of the large overhead caused by error
correction, it is not yet clear if a sufficiently large number 
of qubits can be integrated with each other and if quantum supremacy
can be reached in practical applications \cite{Google}. With this
regard, topological qubits appear more robust than their
superconducting and spin counterparts, owing to the encoding of the
information in their topology \cite{Raussendorf}.

\par To realize a topological qubit, Majorana Fermions, which are
their own anti-particles, are needed \cite{Majorana}. They have been
predicted to exist in solid-state systems such as 1D nanowires (NW)
covered by a $p$-wave superconductor \cite{Kitaev}. Last decade
experiments have mostly focused on demonstrating the presence of
Majorana bound states (MBS) in semiconductor-superconductor (SM-SC)
interfaces, either with InSb or InAs nanowires covered by NbTiN or Al
superconductors \cite{Mourik,Deng,Zhang} or by using Fe 
atomic chains on a Pb superconducting surface \cite{Nadj}. The
presence of MBSs in SM-SC systems follows the so-called ``Majorana
recipe``, which summarizes requirements for the spin-orbit coupling
(SOC) and chemical potential of the semiconductor, superconducting
gap, and Zeeman potential energy induced by external magnetic
fields. Although the $g$-factor of InSb NWs make them particularly
appealing as MBS host, fluctuations in their surfaces can lead to
elusive signatures of Majorana particles.

\par 2D semiconducting monolayers stacked on top of a superconductor, 
as illustrated in Fig.~1(a), offer an attractive alternative to
NWs. Their ultra-narrow thickness allows to reduce the negative
influence of structure variations and lead to stronger proximity
effects with the adjacent superconductor, while still providing the
Majorana conditions \cite{Chu}.  From a device perspective, tuning the
chemical potential of the SM-SC interface can be more easily done with
a 2D semiconductor than a covered nanowire. The operating principle of
2-D SM-SC configurations consists of leveraging the edge bands of
nanoribbons (NR) made of 2-D monolayers with a (significantly) higher
SOC than III-V NWs \cite{Banerjee}, e.g. WSe$_2$. A better detection
of MBS, even with disorder, is then expected \cite{Chu}.

\par Atoms located at the edge of 2D NRs can induce edge states within
the bandgap of the semiconductor. The latter can be used as MBSs,
depending on the SOC strength. Designing such a system requires a high
accuracy that only an \textit{ab initio} simulation environment can
provide. In this work, we therefore present a comprehensive modeling
framework to shed light on  Majorana transport in 2D materials. It
combines density-functional-theory (DFT) and quantum-transport
calculations. The origin of Majorana Fermions is first explained in
WSe$_2$ and PbI$_2$ monolayers that have been selected as testbeds due 
to their large SOC. Finite-size effects coming from the finite
dimensions of the 2D NRs are then discussed by varying the device
length within a fabrication-compatible range. 

\section{Approach}\label{sec:app}
The most important ingredients of the Majorana recipe
\cite{Lutchyn,Oreg}, according to Kitaev's model \cite{Kitaev}, are
listed in Fig.~1(b) for 2D materials. Here, a PbI$_{2}$ NR is chosen as
example because: (i) it possesses an edge band well separated from the
bulk bands and (ii) it shows a strong SOC in the edge band. Figure
1(b) compares a trivial state with Zeeman energy $V_{ZM}=0$ meV (no
magnetic field) to a non-trivial topological one called Majorana
zero-energy obtained at a ``sweet spot`` ($V_{ZM}=0.9895$ meV). It
can be seen that this magnetic field can close the superconducting gap
at the Majorana condition (MC), while increasing its magnitude reopens
it. 

Our \textit{ab initio} framework for Majorana quasi-particles is
summarized in Fig.~2. First, the electronic structure of 2D NRs is
computed with the VASP DFT package \cite{vasp} within the generalized
gradient approximation (GGA) of Perdew, Burke, and Ernzerhof (PBE)
\cite{gga}. Tight-binding-like Hamiltonians are then created by
projecting the plane-wave basis onto a set of maximally localized
Wannier functions. Thereafter, the Tight-binding matrix elements of the $p$ and
$d$ orbitals are grouped to form a basis for the $\vec{L}\cdot
\vec{s}$ SOC operator \cite{Konschuh}. The corresponding SOC 
parameters are chosen in such a way that the DFT bandstructure with
SOC is accurately reproduced. The superconducting proximity effects
are included by applying the Bogoliubov-de Gennes (BdG) transformation
to the Wannier Hamiltonian, i.e. by changing the fermionic operators
($c^{\dagger}=\gamma_1 - i\gamma_2$ and $c=\gamma_1 + i\gamma_2$) to
Majorana quasi-particle ones ($\gamma_{1} = (c^{\dagger}+c)/2$ and
$\gamma_{2} = i(c^{\dagger}-c)/2$). Finally, the external magnetic
field $B$ is accounted for along the transport direction through
Zeeman's field, $V_{ZM}=g\mu_{B}B/2$, and Pauli's spin matrices. The
local density-of-states and transmission function of the finite
systems are calculated with the Non-equilibrium Green's Function
(NEGF) method with the produced BdG Hamiltonian. 

\section{Results}\label{sec:res}
Using the proposed scheme, we study the Majorana
condition in monolayer PbI$_2$ and WSe$_2$ NRs. The edge band of both
materials exhibit significantly different electronic properties. Figure
\ref{fig:3}(a) shows that the
effective mass of the WSe$_2$ (0.7 m$_0$) is smaller than that of
PbI$_2$ (1.8 m$_0$), while the SOC strength in PbI$_2$
($\alpha_{SO}$=0.88 eV\AA, E$_{SO}$=92.5 meV) is considerably larger
than in WSe$_2$ ($\alpha_{SO}$=0.47 eV\AA, E$_{SO}$=10 meV). Figure
\ref{fig:3}(a) further indicates that a semiconductor with a larger $\alpha_{SO}$
has a smaller number of modes near zero energy, making the
detection of Majorana Zero Modes (MZMs) more evident. Moreover, it is shown that the strength of the
superconducting gap induced by the proximity effect ($\Delta$) also
affects the near-zero bands (highlighted in yellow): a smaller
$\Delta$ reduces the density of near-zero energy modes (NZMs).

\par Next, Figs.~\ref{fig:3}(b) and (c) plot the topological
phase diagrams of PbI$_2$ and WSe$_2$ monolayer NRs, respectively. It
can be seen that the considered systems can be brought to a topological
phase (presence of a MZM) from a trivial phase by either tuning the
chemical potential of the 2D semiconductor or by varying the Zeeman
potential. Such topological phase diagrams provide the much needed
guidance to create a topological phase and subsequently a topological
qubit.

\par To confirm the existence of MZMs, we can calculate the
local-density-of-states (LDOS) of the 2-D semiconductors with our
simulation framework. This quantity can be correlated to tunneling
spectroscopy measurements \cite{Zhang}. Figures \ref{fig:4}(a) and (b)
show the energy-resolved LDOS as a function of the Zeeman potential for
PbI$_2$ and WSe$_2$. We observe that the energy gap gradually closes
in the materials as the magnetic field increases before
opening again after the Majorana condition is reached, i.e. at
$V_{ZM}$=0.9895 meV (PbI$_2$) and 0.734 meV (WSe$_2$). The MZMs correspond to
the zero-energy mode peaks in Fig.~\ref{fig:4}(c). 
It should
be noted that the length of the semiconductor NR has been so far
assumed infinite and its electrostatic potential has been kept
constant and homogeneous, thus preventing the investigation of
possible finite-size effects.

\par To go one step further, we now consider the device structure of
Fig.~\ref{fig:5}(a), where the influence of the finite size effect is
introduced through the electrostatic potential depicted in
Fig.~\ref{fig:5}(a) and the generation of NRs made of up to 140,000
atoms. By varying the height of the tunneling barrier, we can assess
the impact of the 2D NR length on the properties of the
MZMs. Figure~\ref{fig:5}(b) shows the energy-resolved transmission 
function with respect to the barrier height $\Phi_{TB}$ and the NR
length $L$, highlighting the MZM and NZMs. Near-zero modes
corresponding to new bound states appear within the superconducting
gap, which is set to $\Delta$=0.1 meV here. As the device length
increases, the number of bound states within the gap grows and the
NZMs move closer to zero energy. In Fig.~\ref{fig:5}(c), the energy
difference between NZMs is extracted as a function of the device
length. From this data, an important parameter called critical length
($L_{crit}$) can be inferred. It provides a condition for the device
length $L$, which should satisfy $L\textgreater\textgreater$ 
$L_{crit}$ to observe a strong MZM signature. 

\par Finite-size effects can be further identified in
Figs.~\ref{fig:6}(a) and (b) where the LDOS is shown for two device
lengths $L$=1.34 and 5.53 $\mu$m, one below and one above
$L_{crit}$=3.3 $\mu$m. We can clearly see the presence of
localized NZMs at both extremities of the device, as expected for MBSs 
\cite{MajoranaReview}. Moreover, the NZMs move closer to the zero
energy when the device length increases, as is also visible in
Fig.~\ref{fig:5}. The MZM and NZMs, plotted for different lengths,
provide another evidence of the localization of these states at the
device extremities. Finally, our results clearly indicate that the
device length should be larger than $L_{crit}$ to obtain a stronger
MZM localization.

\section{Conclusion}\label{sec:conc}
We have developed an \textit{ab initio} simulation framework to
explore the feasibility of topological qubits in nanoribbons made of
2D monolayer semiconductors. Transport calculations with up to
140,000 atoms have been performed in 2D NRs to observe MZM and
topological phases. The Table in Fig.~\ref{fig:6}(c) gives an overview
of the physical settings to create a MZM in WSe$_2$ and PbI$_2$ NRs
and compare them to the same parameters for InSb nanowires. Our
modeling environment opens new avenues for the design of robust
topological qubits with long decoherence times.

\section*{Acknowledgment}
This work was supported the MARVEL National  Centre  of  Competence
in  Research  of  the  Swiss  National Science Foundation (SNSF), by
SNSF under Grant No.~175479 (ABIME), and by a grant from the Swiss
National Supercomputing Centre (CSCS) under Project s876.

\bibliographystyle{IEEEtran}

\newpage

\begin{figure*}
\centering
\includegraphics[width=0.95\linewidth]{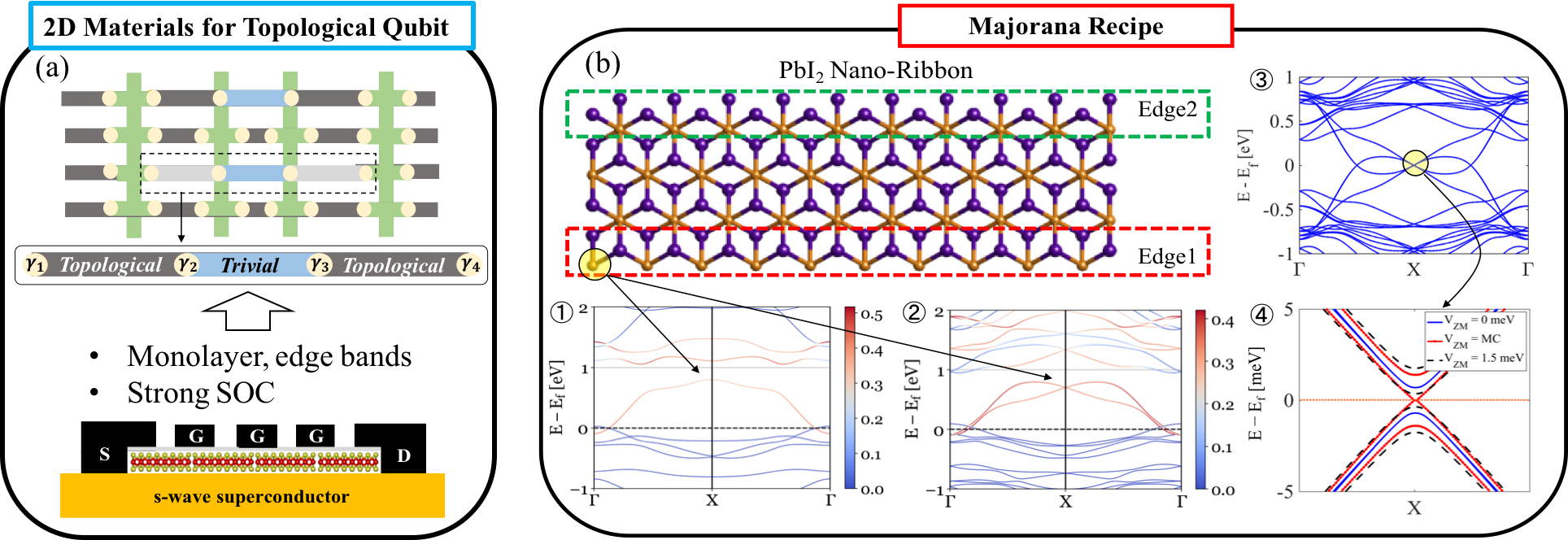}
\caption{Majorana recipe for nano-ribbons (NR) made of 2D
  semiconducting monolayers. (a) Schematic of a PbI$_2$
  NR-superconductor hetero-junction designed to observe Majorana Bound
  States (MBS) from which topological qubits can be created
  \cite{QubitArch}. (b) Atomic structure of a PbI$_2$ NR with a width of
  1.6 nm showing the contribution of the edge atoms to the electronic
  bandstructure computed with VASP without (1) and with spin-orbit
  coupling (2). Sub-plot (3) reports the particle-hole BdG
  bandstructure accounting for proximity effects caused by an $s$-wave
  superconductor stacked on top of the 2D NR. Sub-plot (4) displays
  the band modulation through the Zeeman potential induced by an
  external magnetic field.} 
\label{fig:1}
\end{figure*}

\begin{figure*}
\centering
\includegraphics[width=0.95\linewidth]{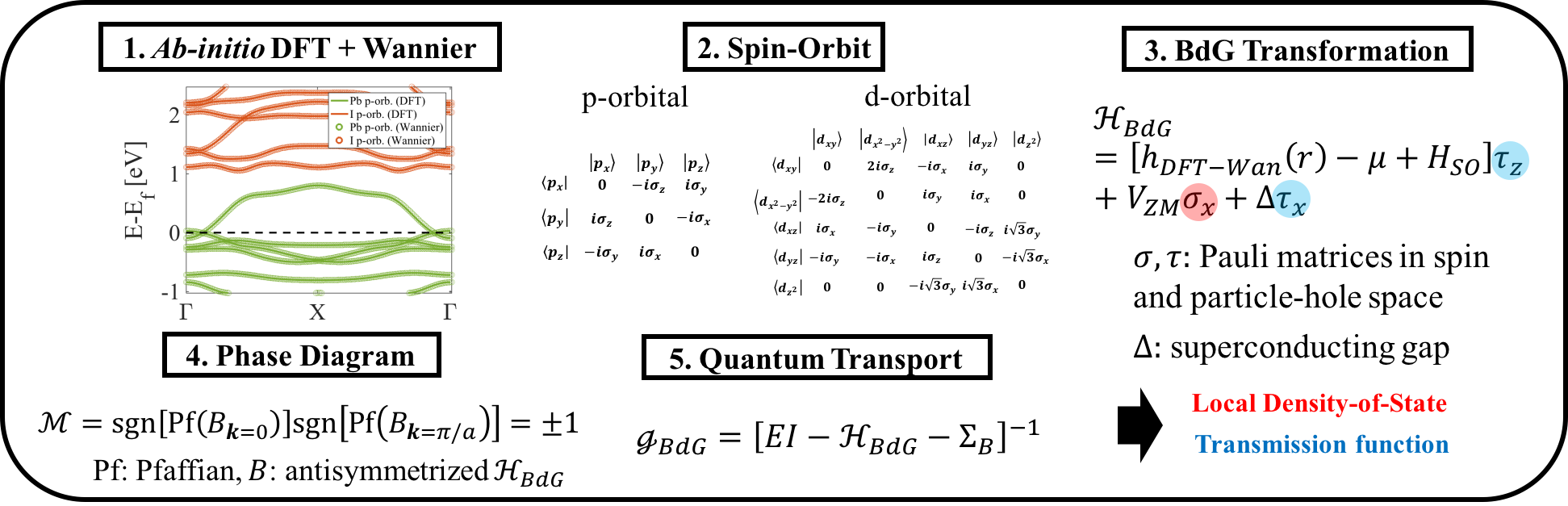}
\caption{Summary of the proposed \textit{ab initio} simulation
  framework. (1) The electronic bandstructure obtained from
  density-function-theory (DFT) is first projected to a Wannier
  function basis (here: PbI$_2$). (2) Inclusion of SOC in the Wannier
  Hamiltonian to reproduce the DFT bandstructure with SOC. (3) BdG
  transformation of the Wannier Hamiltonian to capture the influence
  of the superconductor stacked on top of the 2D material. (4)
  Determination of the topological phase with the help of Pfaffian
  invariants. (5) NEGF-based quantum transport calculations using the
  BdG Hamiltonian.}
\label{fig:2}
\end{figure*}

\begin{figure*}
\centering
\includegraphics[width=\linewidth]{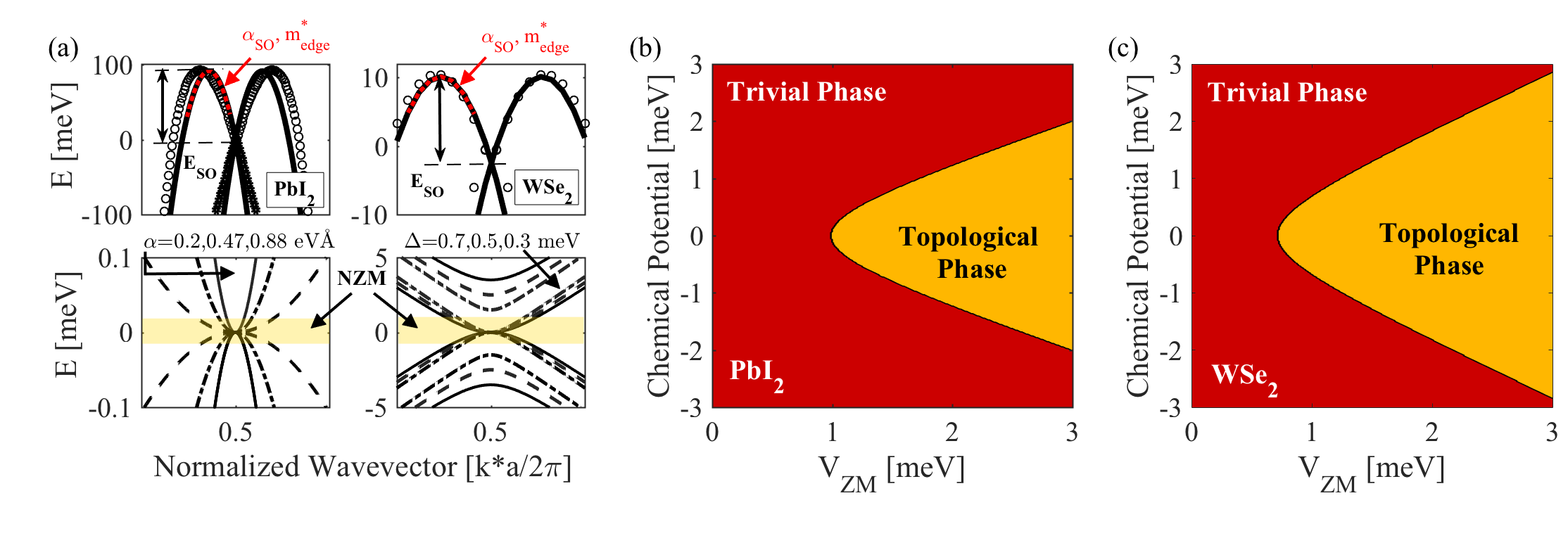}
\caption{(a,top) Extraction of the effective mass of the edge band
  (m$_{edge}^{*}$), spin-orbit energy (E$_{SO}$), and spin-orbit
  coupling strength ($\alpha_{SO}$) from the DFT bandstructure with
  SOC (symbols). The solid lines represent the fit obtained with a 1D
  Majorana model \cite{Oreg}. (a,bottom) Effect of $\alpha_{SO}$ and
  the superconducting gap ($\Delta$) on topological states at the Majorana condition. Here, NZM refer to near-zero energy modes, which
  are the modes in the highlighted region. (b) Topological phase
  diagram ($x$-axis: Zeeman potential energy, $y$-axis: chemical
  potential of the material) of PbI$_2$. (c) Same as (b), but for
  WSe$_2$.}
\label{fig:3}
\end{figure*}

\newpage

\begin{figure*}
\centering
\includegraphics[width=0.95\linewidth]{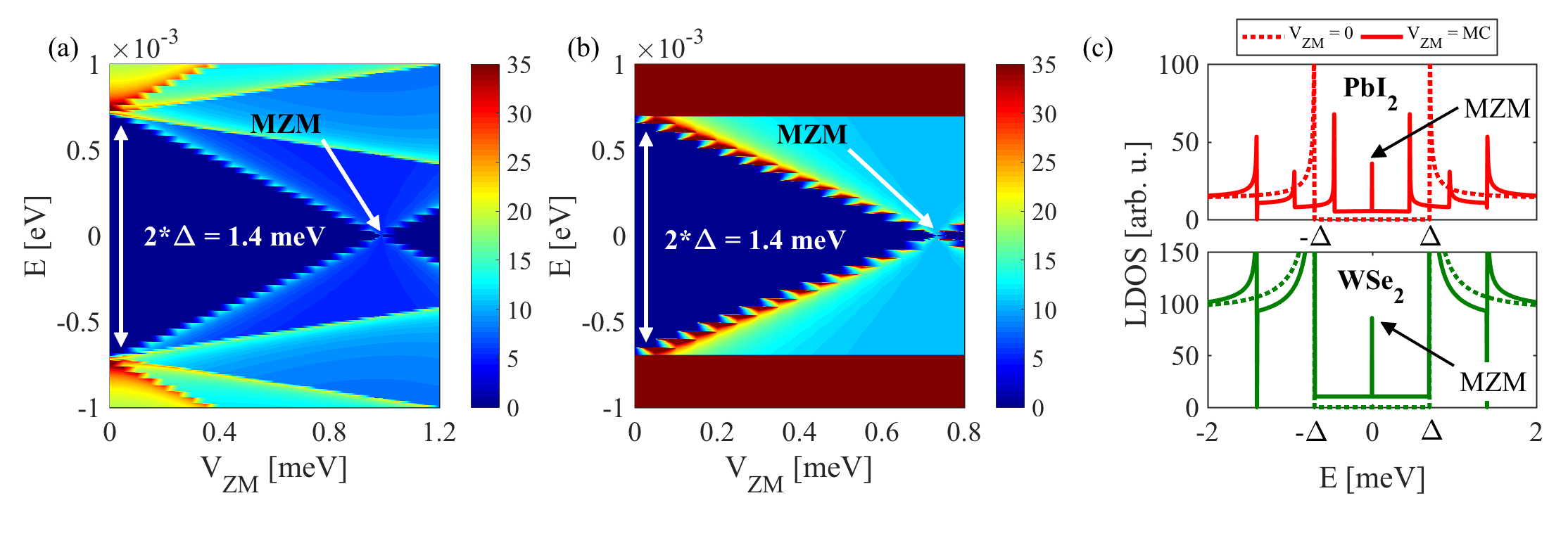}
\caption{(a) Energy-resolved local-density-of-states (LDOS) of PbI$_2$
  as a function of the Zeeman potential energy $V_{ZM}$. (b) Same as
  (a), but for WSe$_2$. (c) Same as (a) and (b), but for $V_{ZM}$=0
  meV (dahsed line) and for the $V_{ZM}$ providing the Majorana
  condition (MC, solid lines). A superconducting gap $\Delta$ = 0.7
  meV is assumed. The MZM peaks are clearly visible when $V_{ZM}$ is
  tuned to the Majorana condition (MC), where $V_{ZM}=g\mu_{B}B/2$, $g$: land\'e $g$-factor, $\mu_B$: magnetic moment, B: external magnetic field.}
\label{fig:4}
\end{figure*}

\begin{figure*}
\centering
\includegraphics[width=0.95\linewidth]{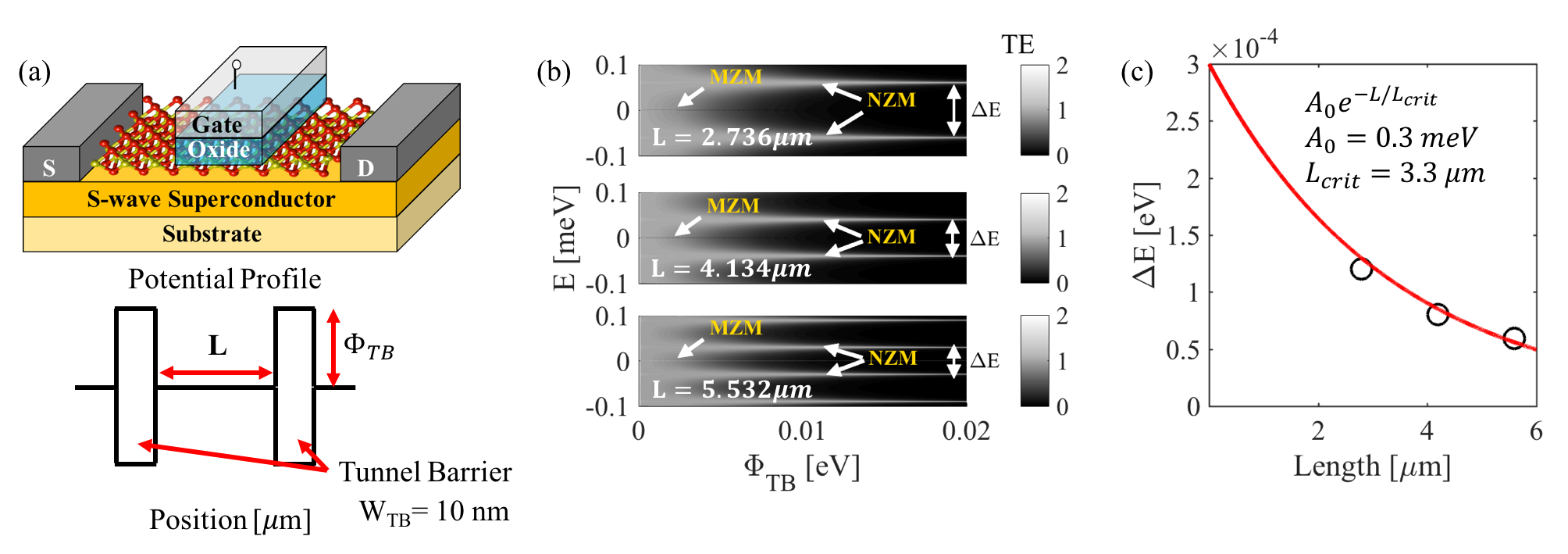}
\caption{Finite length effects in PbI$_2$ NRs. (a) Device structure
  and electrostatic potential profile used to model finite-size
  effects in 2D NR with tunneling barriers. The barrier width is kept
  fixed at 10 nm, while the barrier height ($\Phi_{TB}$) is varied
  between 0 and 20 meV to observe finite-size effects. (b)
  Energy-resolved transmission functions vs. barrier height at
  different device lengths $L$. (c) Extraction of the critical length
  $L_{crit}$ from the energy difference between near-zero energy modes
  ($\Delta$E) as a function of the device length $L$.}
\label{fig:5}
\end{figure*}

\begin{figure*}
\begin{minipage}{0.58\linewidth}
\centering
\includegraphics[width=\linewidth]{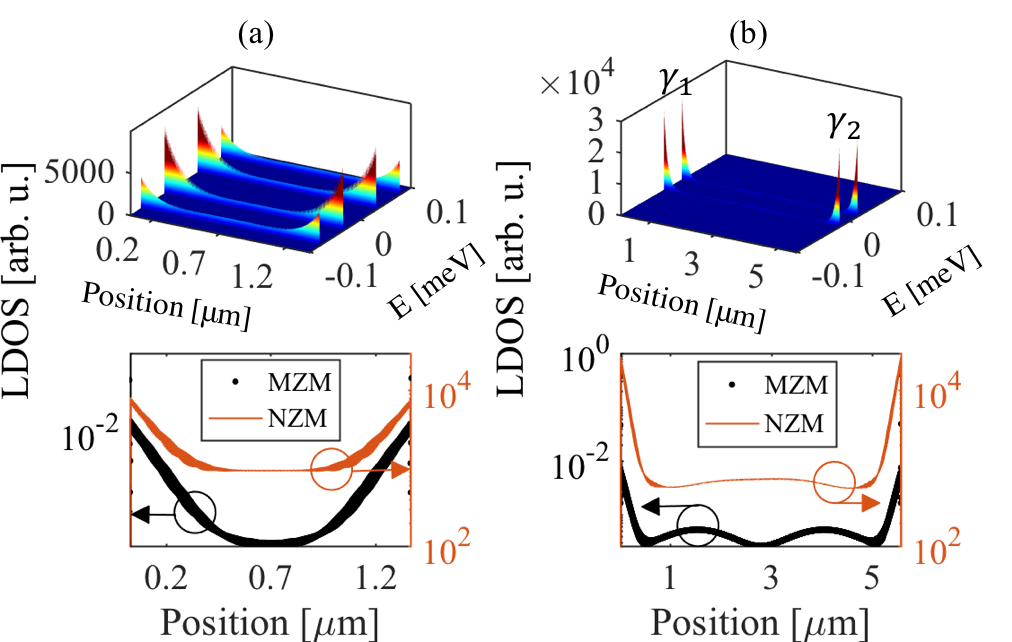}
\end{minipage}
\hspace{0.1cm}
\begin{minipage}[b]{0.33\linewidth}
\begin{tabular}{|c|c|c|c|c|}
\hline
(c)&PbI$_2$&WSe$_2$&InSb \cite{MajoranaReview}\\
\hline
Structure & NR & NR & NW\\
MZM transport & Edge & Edge & Bulk\\
E$_{SO}$ [meV]&92.5&10&0.05 - 1\\
$\alpha_{SO}$ [eV\AA]&0.88&0.47&0.2 - 1\\
$\lambda_{SO}$ [nm]& 0.48 & 2.3 &230 - 50\\
$L_{crit}$ [$\mu$m]& 3-4 & -- & 1-2\\
$\lambda_{MFP}$ [nm]& --  & -- & 200-300\\
Width [nm] & 1.6 & 2 & 80-100\\
Thickness [nm] & 0.68 & 0.65 & 80-100\\
Proximity effect & High & High & Low\\
$m^{*}$ $[m_0]$&1.8&0.7&0.014\\
$g$-factor& -- & -- &40 - 50\\
\hline
\end{tabular}
\end{minipage}
\caption{(a,top) Energy- and position-resolved LDOS in a PbI$_2$ device
  of length $L$=1.34 $\mu$m. (a,bottom) Zero-energy and
  near-zero energy mode as a function of the position. The
  localization of these states at the two ends of the device is
  clearly visible. (b) Same as (a), but for $L$=5.53 $\mu$m. (c)
  Comparison of the required physical settings in WSe$_2$ and PbI$_2$
  nanoribbons and in InSb nanowires to build a topological qubit based
  on a Majorana bound state.} 
\label{fig:6}
\end{figure*}

\end{document}